# Embracing ambivalence in studying technology acceptance: A qualitative study on automated visual software for live music performance

Spagnolli A, Mora D, Fanchin M, Orso V, Gamberini L

Dept. of General Psychology and HIT research centre, University of Padova, Italy

**Abstract (147)**

Although the live music entertainment sector does not directly fuel the current debate on automation, it might harbor positions that resonate with it. In this paper we study a prototype software application helping DJs and VJs to accurately manage and even automate the synchronization of visuals with music during amateur or professional live performance. The goal of the study was to unravel VJs' and DJs' ambivalent positions about this software. We preliminarily investigated VJs' and DJs' perception of their sector of activity with seven face-to-face interviews and an online survey (N = 102); then, we asked DJs and VJs (N = 25) for their opinions about our prototype software application. Four core controversies were identified in their answers, along with a set of arguments mobilized to take side on them. The advantages of focusing on ambivalence and argumentation when studying users' response to new media are discussed.

**Keywords:** ambivalence, qualitative, automation, live music performance.

1. Introduction

The market success of a new technological device and its adoption by users does not depend on its technical properties alone. The agenda, habits, and relations of its potential users, as well as the nature of its infrastructure or the subtle working of political dynamics, shape the actual profile by which a technical device is eventually recognized and received (Gagliardi, 1990; Kling, 1980). The possibility that a device is resisted against and eventually dismissed despite its technical merits is a well-known phenomenon that investors hope to prevent and scholars try to understand. Currently, automation based on artificial intelligence, sensing technologies and networked infrastructure is one technology whose eventual physiognomy is greatly uncertain. Its reception hovers between hope and fear, and it is alternately perceived to empower and to deskill, to help and to deceit (e.g., Zhang, Dafoe, 2019). Although such debate mainly refers to application domains such as transportation (automated cars, Shariff, Bonnefon, Rahwan, 2017), manufacture (industrial robots, COMEST, 2017), law (predictive justice, Angwin, et al, 2016), and health (AI diagnostics, e.g. Nature, 2018), as well as to recommendation/personalization services for media fruition (Klinger, Svensson, 2018), we found as much uncertainty when studying automation in the sector of live music performance.

Live music performance includes events varying from weekend nights in local pubs to large music festivals with an international audience. Computerized automation can assist in several ways during such events: to simultaneously shift dozens of faders at pre-defined values, to swiftly change tempo while mixing different music tracks in sequence, or to finely synchronize visuals with music, adding a pictorial accompaniment to the soundscape. This last aspect is the one managed by the prototype application that we studied, a VJ application that allows to modify the appearance or movement of selected elements of a visual animation in synchrony with music, for example changing the diameter of a spiraling circle in accordance with the music pace. The content selection, the settings and the progress of the visual show can be defined by the human user or automatically managed by the software in a sort of autopilot mode. The professional figures expected to use this software are live performers (i.e., VJs and DJs) who use that content to create, edit and play visuals during live events.

As soon as we started to collect the prospective users' opinion of the software and tried to make sense of the answers collected, we were struck by the pervasive ambiguity of their position, epitomized by statements such as: "This could be an advantage but often it is badly exploited."(D1) or "I think it is useful but also risky" (VJ3). Instead of methodologically neutralizing ambivalence by forcing respondents to take one side, we decided to acknowledge the ambivalence generated by the topic of our investigation and to make it the focus of our analysis. In the rest of the paper we will describe the qualitative approach we followed to do so; we will start in the next section by briefly introducing the concept of attitudinal ambivalence.

**2. Ambivalence: classic and argumentative approaches**

Opinions are considered ambivalent when both positive and negative positions are endorsed about the same topic. Although attitudinal ambivalence has been given some attention since Kaplan's seminal work (1972), most research on user's decision to adopt a given technology does not consider ambivalence. The Technology Acceptance Model (Davis, et al., 1989; Lee, et al. 2003; King and He, 2006; Yousafzai et al., 2007a,b; Williams et al., 2015, Venkatesh and Bala, 2008), which catalyzes most works on technology acceptance (Taherdoost, 2018), measures the intention to purchase or adopt a technology as one value on a continuum, without space for ambivalence. The measure itself is collected via questionnaire items whose formulation remains fairly unaltered across devices and domains thereby making it difficult to voice specific reservations and concerns that would be of great value to designers (Hornbæk and Hertzum, 2017).

Attempts at capturing users' ambivalence, although the term itself might not be used explicitly, are made by adopting qualitative approaches. Participants are involved in face-to-face interviews to explain why they used a given technology on a daily basis or why they stopped doing so (Alapetite, et al, 2009; BenMessaoud, et al, 2011; Heikkilä and Smale, 2011; Hennington et al, 2009; Mahzan and Lymer, 2008; McNaney et al, 2014; Middlemass et al, 2017; Nakatani et al, 2012; Nguyen, 2017; Ouadahi, 2008; Peek, 2016; Prior, 2018; Rahim, 2008; Stock and Merkle, 2017; Weidinger, 2017; Widuri, 2017). Or they can be gathered in workshops and focus groups to define the functions that are mostly used (Brinkel, 2017; Mallat, 2007; Nguyen, 2017; Peng, 2016) or invited to explain in writing what would make them more supportive of the technology at stake (Alapetite, 2009; Prior, 2018). Sometimes, participants are observed while using the device under investigation (Heerink, 2009; Salmona and Kaczynski, 2016). Once qualitative data is collected, the analysis proceeds bottom-up to identify which elements would facilitate the acceptance of a given technology and which elements would decrease it. This way of proceeding exemplifies the approach by 'tradeoffs' (Rosson, Carroll, 2002) or 'tensions' (Tatar, 2007) in which pros or cons of some design options are listed.

What is still missing from this approach is the acknowledgment of the rhetorical, argumentative level at which users act when they express their opinions. Users re-elaborate the interviewers' questions (see Schwartz, 2007; Suchman and Jordan, 1990) within a "locally coherent versions of the social and moral world" (Potter, 1998, p. 244) in which taking a position means to reject other alternatives (Billig, 1996; Potter and Wetherell, 1987; Wetherell and Potter, 1988) and to be attributed a consequent social identity (Antaki, 1992). At this level, controversies, dilemmas and conflicts are not blots to be obliterated from the data; they represent the substance of the users' answers, mapping the complexity of the issues for them, and the intricacy of their implications. By analyzing the argumentative, rhetorical meaning of the users' opinions, the *many* controversies gathering around a given topic can be unpacked and the *arguments* responsible for supporting the different sides in such controversies can be identified (the 'ideological dilemmas' of everyday talk, Billig et al, 1988). In this way, instead of finishing the analysis with a list of software features that are either appreciated or hideous, we will also know the meaning they have to the user and the reasons why they are seen either positively or negatively by them. In the next section we will describe more in details the method followed in our study to understand the argumentative context framing users' expressed opinions of VJ software.

## 3. Method

In order to grasp the full meaning of the users' arguments, one must first share with them some background knowledge to get an understanding of what is changing, valuable and difficult in their sector of activity. Therefore, we started our study by interviewing a number of users face-to-face and a larger number online to gather their view of the live music performance sector. The second phase of the study, instead, involved a new set of DJs and VJs, who were interviewed about their opinion of VJ software. In the rest of this section, the sample and data collection methods used in these two phases of the study are described.

**3.1 Phase 1: Users' perception of their sector of activity**

*Interviews.* A small number of stakeholders (four VJs, one DJ, one laser show designer) and one venue owner were interviewed face-to-face. They were asked their opinion about what makes one a respectable professional in their sector, the main difficulties in doing a good job in their sector, the major changes over the last few years and when a visual/music performance is high-level. Follow up questions were added if needed, in a semi-structured interview format. The interviews were videorecorded after receiving signed authorization by participants, and later transcribed. Interviewees' age was 38.2 on average (SD = 5.9) and they were all men (reflecting a gender imbalance typical of this sector, Larsson, 2017). They declared to have all been involved in live music performance in clubs for an average of 14.4 year (DS = 7.1) and many had the experience of performing in music festivals as well (86%). In terms of nationality, most were Italians (5 out of 7), one was German and one Portuguese.

*On-line survey.* An online survey was administered via a commercial online survey platform (SurveyMonkey); it included the same open questions as the interview: What makes you a respectable professional in your sector?, What are the main difficulties in doing a good job in this sector today?, Which major changes have you noticed in your sector of activity over the last years?, When is a visual performance very high-level?. Some closed questions about software requirements were also included, but they are not of interest here. The survey ran online in June and July 2017. After cleansing the dataset from respondents providing inappropriate answers (N = 5) or from respondents who did not fulfill the inclusion criteria (i.e., they were not performers nor owners/promoters, N = 18), the sample included 102 respondents, aged 33.18 years on average (SD

= 7.9), 79 of which were men. About one half of them were Italian, but the sample comprised respondents from the USA, Asia, South America, Australia and several European countries. Respondents declared to perform all over the globe, except for Africa, and to be involved in live performance in clubs (85.3%), music halls (38.2%) and/or festivals (55.9%). Of the 102 respondents, 73 were DJs or VJs; 48 of them declared to personally play visuals live and 14 to create their own visual content themselves.

*Data analysis.* Regarding the interview data, three members of the research team read the seven interviews together and extracted four recurring themes. Regarding the online survey, whose sample was larger than the interviews', we proceeded by first defining a coding scheme, and then by having two members of the research team independently code all answers (Table 1). Disagreements between them were solved jointly. The themes emerged from this phase are reported in section 4.1: the sense of a growth in visuals demand, pressing budget issues for VJs, the notion of live performance as art and a large ambivalence towards VJ software

Table 1. Coding categories and intercoders' agreement.

| Questions | Categories | Intercoders' agreement |
| --- | --- | --- |
| What makes a respectable professional? | Quality, creativity, experience, passion, equipment | 91.84 |
| Main difficulties in the sector? | Cost, speed, low cooperation, technical issues, rapid updates, big competitors, poor acknowledgment of their role, regulations, market saturation | 97.96 |
| Main changes over last years in the sector? | Digitalization, social media, budget, visuals, music taste | 92.86 |
| What makes a high-level | Perfect synchronization, originality, advanced | 85.8 |

| visual performance? | technology | |

## 3.2 Phase 2: Users' position on VJ software

The interviews carried out in the second phase of the study aimed at collecting VJs and DJs positions about VJ software. The interviews dwelled on software characteristics such as automation and intuitiveness that were likely to clash with some defining aspects of the interviewees' activity, such as originality and skills. Questions directed to VJs and DJs differed, to fit the different role VJ software has in their activity. The main questions to VJs were the following:

- Does software influence creativity positively or negatively?
- Would software that automatically produces high-quality visuals shrink or enlarge the VJ market?
- Would easy-to-use VJ software allow high artistic results?

DJs were asked instead:

- Does software influence creativity positively or negatively?
- Would you prefer to work with a VJ or to have high quality visuals automatically synchronized with your music by some software?
- Some VJs do not appreciate automatic visual software, because they believe it would lower the artistic level in their sector. Do you think they are right?

Follow-up questions were then added to the three main questions listed above in order to clarify the interviewees' answer. 25 artists were contacted, 12 VJs (2 women and 10 man) and 13 DJs (1 woman and 12 men), all Italian. None of them participated in the first phase. The mean age of the VJ's sample was 35.25 (SD = 4.55) while the mean age of the DJ's sample was 32.92 (SD = 6.96).

VJs declared 11.91 years of activity on average (SD= 3.42) while DJs declared 13.23 years of activity on average (SD = 5.29).

These interviews were conducted remotely using a voice message application (WhatsApp) on mobile phones. The interaction was synchronous: the interviewer recorded a question as a vocal message and sent it to the interviewee, who immediately replied in the same way. The informed consent was instead collected prior to the interview via email. The interviews were transcribed and analyzed to identify the users' positions. More than understanding whether such positions were favorable or unfavorable, we were interested in the content of each position and in the arguments mobilized to take it. Thus interviews were parsed to identify controversies and arguments. Let's consider for instance the following passage:

> Interviewer: Do you think that easy-to-use VJ software allows to achieve high artistic results?
>
> Interviewee (VJ13): I think that what matters is the result, not the praxis. If we are talking about visual arts, then what matters is the final visual result. To consider that the hands are the only point of artistic value is a little of a 20th century perspective...

In his answer, the interviewee takes a favorable position about using software in artistic work. Such position is unexpressed (Van Eemeren, Grootendorst, 2016); it is taken by opposing a counter-position ("considering that the hands are the only point of artistic value") and is supported by two arguments, "what matters is the result" and "considering that the hands are the only point of artistic value is a little of a 20th century perspective". This passage, therefore, evokes one controversy, namely whether using software can result in artistic work or not, and mobilized two arguments to support one side in such a controversy.

Once this analysis was completed on all interviews, controversies and related arguments were grouped by similarity. The results are reported in section 4.2.

**3.3 Ethics**

The research was carried out in compliance with the Code of Ethics of the American Psychological Association and with the European data protection law (Directive 95 /46/EC, 2002 /58/EC and, since May 25, GDPR 2016/679). The goal of the data collection was disclosed in the informed consents ("understanding VJs and DJs preferences in terms of software to prepare and manage live music performance" or "the relation between software and creativity"), along with the general purpose of the project, the sponsoring institution, the contact information of the research team and the type of data collected. No deceit was used. In the face-to-face interviews the consent form was printed on paper and signed before starting the interview; in the online survey, respondents who did not accept the terms described in the front page could not proceed with the rest of the survey; in the WhatsApp interviews the copyright form was sent via email and returned via the same channel prior to the interview. To the seven interviewees met in the first phase, a beta version of the software was promised as a compensation for their participation and time; no compensation was given to the other participants. The demographic data collected (age, gender, nationality, performance region, type of performance events, role in the event, years of experience) is justified by the research goals in obedience to the data minimization principle. Participation was voluntary and the possibility to withdraw was given at any time.

**4. Results**

We first describe the users' perception of the sector of activity, and then the main controversies regarding visual software.

**4.1 Users' perception of the live performance sector**

*Visuals' demand is growing.* Respondents see the subsector of visual performance as growing. They declared that "[there is] a growing interest in lights and visuals" (F-VJ4, F-VJ5); "there are plenty of musicians who are interested" (F-VJ5); "I don't see there are really competitors. Not in a problematic way. So it's still a small field. There are only a few, I would say, which are well-known and so it's more about making the scene even bigger, being open". Consistently, when asked about the main difficulties in doing a good job in their sector, VJs tend to mention market saturation less often than DJs or venue owners (Fig 1).

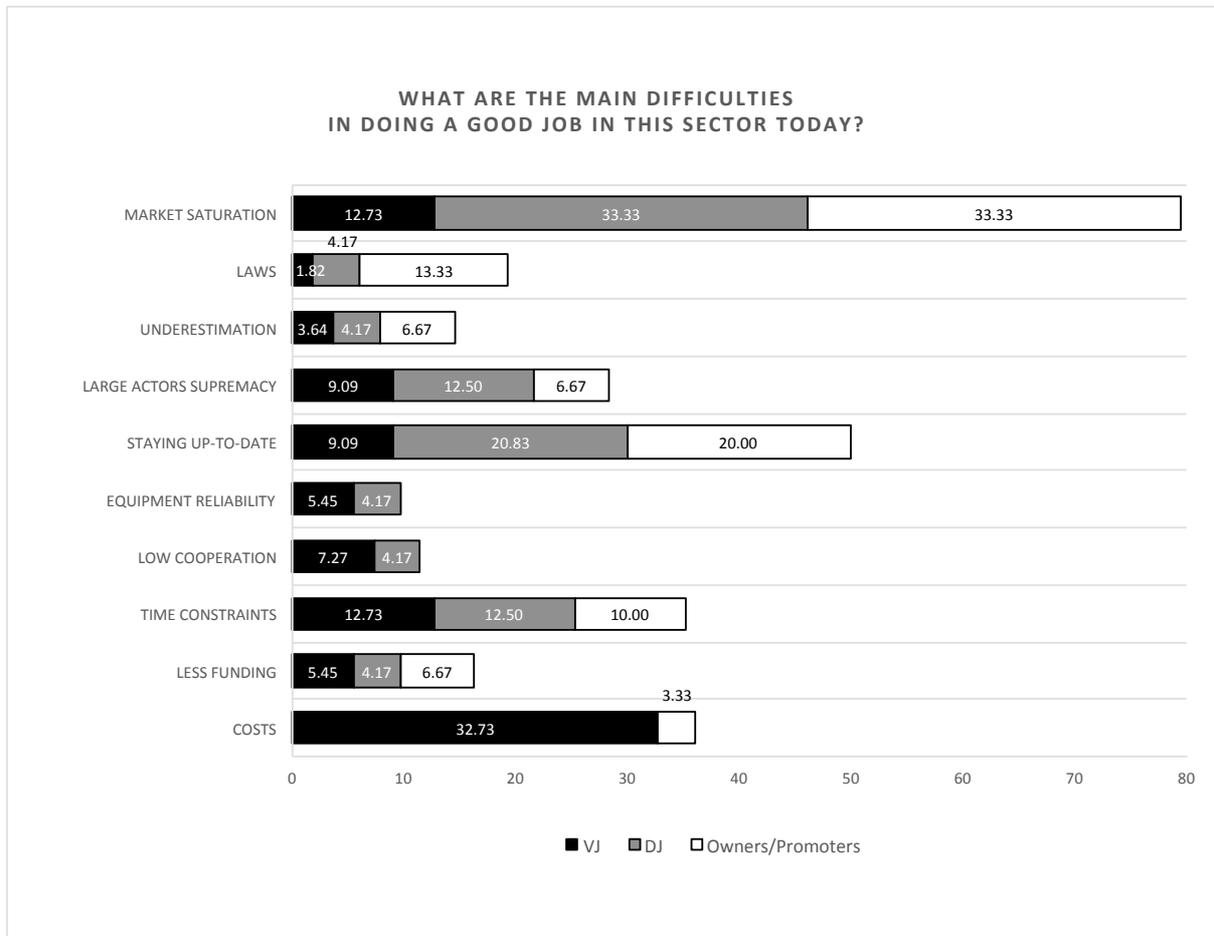

**Figure 1.** Answer categories and percentage frequency (N = 102), broken down by category of respondent (VJs, DJs, Owners/Promoters); answer categories are not mutually exclusive

*Budget issues.* VJs tend to mention budget difficulties (Figure 1); more specifically, they complain more than DJs about the equipment-related costs. The reason is that whereas DJs only need to invest on music, VJs need to personally invest on visual editing software and equipment, which is rarely present at the event venue: "for a young person starting this activity the main obstacle is surely the cost of the software" (F-VJ2); "I bring my laptop, I bring my adapter, and my controller." (F-VJ5, F-VJ6, F-VJ1).

*Live performance is artistic work*. When asked "what makes you a respectable professional in the live performance sector" (Figure 2), the online survey respondents mentioned precision of the performance (quality), creating an original performance, experimenting with new techniques and keeping updated about the latest musical trends (creativity), possessing the skills deriving from many years of activity (expertise), having a high interest, determination and devotion to their work (passion) and skilfully using high quality hardware and software equipment (equipment). These answers confirm the importance of equipment already mentioned above. They also show that DJs and VJs see themselves as *artists*: being a good VJ or DJ does not only consist of entertaining the audience, but also of making their musical or visual performance unique: "I only use things that I did myself. I do not use things other people did or that are downloaded from the Internet. It is important to be original, the beautiful part of our job is precisely to create"(F-VJ1); "it's very related to artistic thinking (...) uniqueness is essential in my job" (F-VJ4); "my job is to create something which is unique, which not anyone has" (F-VJ5).

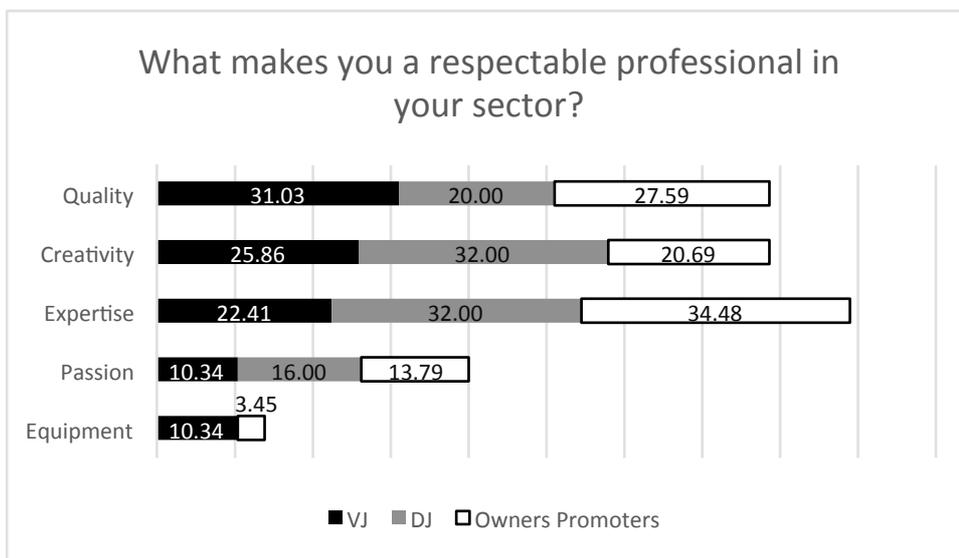

**Figure 2.** Percentage frequency of respondents mentioning each characteristic, broken down by role (N = 102). Characteristics are not mutually exclusive.

*VJ software is good or is it?* On the one hand, digitalization is spontaneously mentioned in the survey as the most noticeable change in the live music performance sector over the last years by all VJs and 45.5% DJs. The interviews also suggest a harsh need for better software interfaces: "video technology and the visuals are more recent technology (light and sound). We haven't sophisticated equipment"(F-VJ4); "the weakness of the software is that it is still young; it has a modern approach but still lots of things to implement" (F-VJ3). On the other hand, respondents fear that software can lower the artistic quality of a performance: "even who is not a designer can use this software and improvise because the software comes with a set of pre-made effects" (F- VJ3); "in my opinion, marketwise, my main competitor is the ease with which you nowadays can be a VJ" (F-VJ2).

In synthesis, users perceive live visuals as a growing sub-sector, characterized by digitalization and equipment-related costs. Live music performance is an artistic job, and adhering to artistic standards qualifies one as a good professional in such job. Being competitive in a saturated market is important to DJs.

**4.2 Controversies about software for live music performance**

When expressing their opinions about VJ software, our interviewees evoked four main controversies. We will describe them one by one, along with the arguments mobilized to support the side taken. We would like to clarify that the frequency with which each argument recurs in the answers is reported in the text for the sake of accuracy. However, the relevance of the arguments depends on being part of the repertoire of arguments spontaneously used in this context, not on the frequency of their occurrence.

*First controversy: Positive/negative impact of software on creativity*

This controversy concerns the impact that using VJ software can have on creativity and art, whether positive or negative. To take a side in this controversy, interviewees mobilized the following arguments:

- *Technology is entrenched in VJs' job; artists must keep up-to-date with it.* Software cannot be dispensed of, especially by VJs: "our sector is connected 100% to computer and software" (WVJ8), "or else I should be doing shadow play" (WVJ12). This argument in present in 10 different interviews and seems to work as a disclaimer, preventing possible misunderstanding coming from subsequent talk.

- *Technologies do not determine the final product of the artistic process: they only provide resources.* Software in itself has no positive or negative consequences: it puts some resources at the service of the artists' creativity and can augment it. This argument is used by 11 interviewees, for instance: "software empowers and opens new ways, simplifying some steps or discovering creative veins that were not possible without" (WVJ3); "software allows to create a more precise and possibly elaborated performance at no detriment of its artistic value"(WDJ5).

- *Artists are responsible for using tools creatively or not.* Eight interviewees use the argument that artists should actively study which features of the software can better fit their needs and serve their creativity. This argument simultaneously relieves the software from being attributed any intrinsic negative power, and sets a criterion to distinguish between good professionals and bad professionals: "when one gets some software, one should start studying it and checking what it allows you to do so you can use it at its best and this is necessary to be creative, first understanding its functionalities and then focusing on those that better express your ideas" (WDJ2).

- *Technology interlopes between the artist and its audience*: one DJ in our sample, who positions himself in the negative side of the controversy, declares to be "unfavorable to software; I have always used vinyl, and been in touch with the audience, speaking to it in a microphone and make it have fun" (WDJ9).

*Second controversy: Automation can/cannot result in an artistic product*

This controversy regards whether a performance could still be artistic if much of its operations are carried out automatically by some software. To take a side in this controversy, interviewees mobilized the following arguments:

- *There is art as long as there is some human intervention.* In order for a performance to be artistic, it needs some meaningful human intervention. This does not prevent an artist to use software, but it prevents products of pure automation to be called artistic. The argument is found in 12 interviews, for example: "I'd say that unless I operate in it with my idea, no software can produce any high quality graphics"; "when I attend a VJ set I notice whether the VJ is an artist or is merely a technician who is playing the downloaded content and working on it with solutions readily offered by the software" (VJ11); "all artistic forms need improvisation and here is where you see how good the artist is" (WDJ4); "only the artist can empathize; a product might be beautiful but surely the human artistic stamp is missing (WDJ1); "[a human artist] makes a situation more real, including by doing mistakes, which make everything look truer" (WDJ12); "no software has taste, taste is what differentiates a DJ from another (WDJ13)". In four interviews it is underlined that "when the software becomes a machine that does not need any human (...), this would be a limit for VJs (WVJ10);" "yes I side with VJs who say that a totally automated software would trivialize them as artists and replace them"(DJ7).
- *Being a beginner or having a very low budget are extenuating circumstances.* Five interviewees accept that beginners use software more slavishly, sticking to downloaded content and predefined effects, because they are still practicing their own skills and need some reliable backup during public performances: "at the beginning of my career I used some automatic tools that could generate visuals almost by themselves" (VJ9). It is also expected that low-budget events do not invest much on art: "this (software) could bring more visuals in situation that otherwise could not afford them" (VJ8)

- *Technical mastery is not enough for being an artist.* Six VJs argue that technical skills do not make one an artist. This comes in reply to a question about intuitive software, but is of relevance to this controversy since it supports the position that technical mastery per se is not the essence of artist work: "ease of use is not what matters, what one wants to represent and how one uses some tools to achieve that representation, this is what counts"(WVJ8); "When VJing, ease of use does not count, what counts is the quality of the content and the ability to mix it, to modify it in real time"(WVJ2).

- *A good artist is one who can master the technicalities of his/her own tools.* In contrast with the previous argument, one participant argues that technical abilities are so core to their professional identity that any sort of automation would disqualify an artist: "the DJ is a person who plays records at the right pace and makes the shift from one piece to another seamless, which unfortunately today is automatically managed by some software"(WDJ7).

*Third controversy: Intuitive software is professional/for beginners*

This controversy has to do with the level of technical expertise and in particular the kind of computer skills a true artist is expected to master. The question is whether a serious, respectable artist in this sector is supposed to be able to deal with complex interfaces and even programming language, so that using software with an intuitive interface would look unprofessional. To take a side in this controversy, the following arguments were mobilized:

- *Intuitive interface must not mean limited.* Intuitive software needs not to be a synonym of poor software, namely software which only offers some basic features. For four interviewees: "it would be limiting if an easy interface is combined with poor functionalities, so to speak, for instance in the possibility to manipulate the audio signal or in the generative functions library"(WVJ10); "sometimes the most user-friendly software is also the most commercial software, and then the one that constraints me more" (WVJ5).

- *Intuitiveness frees time for real artistic effort* Intuitive software frees the artist's time from mechanical work to focus on more artistic aspects of it. This is mentioned in eight interviews "software reduces time waste due to the most technical parts of mixing and allows a DJ to focus more on music selection" (WDJ8); "If I am playing music somewhere (…) and let the records be automatically synchronized in BPM [beats per minute], why would I do that? To focus on other things (...), for instance on live remixing or sound equalization so that my performance improves"(WDJ2); "software that is say intuitive, that has the buttons in the right place, that allows one to work on it and interact with what one does without opening too many pages, this kind of software is helpful" (WVJ6).

*Fourth controversy: VJ software shrinks/boosts VJs job market*

A core controversy related to using VJ software regards the repercussion on the job market, especially if the software has some degree of automation and can take care of some tasks on behalf of the artist. There are many categories of artists whose job might be affected by software of this sort, and their interests might be in conflict. The category mostly at risk is VJs', since automatic software can replace them in a profession that is already seen as ancillary to DJs. To take a side in this controversy, interviewees mobilized the following arguments:

- *Automation is more reliable than humans.* Some DJs would prefer software to manage visuals in their live performance instead of a VJ, because it allows to keep a higher control on the final product: "software would allow me to create my things without explaining all my ideas to another person, which is always quite difficult (WDJ6)"; "I am safer if I can rely on myself only because I know all deadlines (…) to work with another person might be unreliable sometimes" (WDJ1).

- *Market will evolve, not disappear.* VJs' work might change instead of disappearing "automatized graphics need to be created by somebody and, in order for them to be good, you need a good VJ; so I do not think that the market will shrink" (WVJ4). In four interviews it is mentioned that automated software will layer the market: automation will satisfy low budget venues, but there will be plenty of situations in which an artistic touch will be preferred: "the VJ moves from working in clubs only to producing these programs."(WVJ4); "this software allows the artist to work not only in clubs but also in artist installations or dance performances" (VJ12).

*Affordable software is a Trojan horse*. One optimist position is that the job market will eventually grow: automated software will help visuals be more pervasive and customary, and this will increase the demand for more artistic content. This argument is mentioned in six interviews: "once you realize that you cannot go beyond what the software itself can already offer then a demand grows for new content, a demand that did not exist before" (WVJ8); "there is only one phase in which the market shrinks where good VJs will suffer; then the aesthetics will be saturated by the new software solutions and it will develop into something different; this phenomenon is cyclic, it is nothing new" (WVJ5); "with an automated software visuals could be affordable to everybody and much more venues could invest in it and have visuals every night." (WDJ7).

## 5. Discussion and conclusions

We synthesized in four controversies all disputable points evoked by our interviewees while expressing their opinions about VJ software. VJs and DJs touched upon whether software helps or hinders creativity, whether automation and art are compatible or at odds, whether intuitive software is adequate for skilled artists or not, and whether VJ software will shrink or enlarge the VJs' job market. The first phase of our study helped depict the large scenario of these controversies, in which digitalization is perceived as one of the biggest recent changes in the live music performance sector, staying up-to-date is the second most important difficulty in VJs' and DJs' job and conflicting demands impinge on different kinds of artist. The specific arguments mobilized by the interviewees to take side in such controversies and identified in the second phase of the study represent the main findings of the approach to acceptance adopted in the present work: they sort out many apparent contradictions in the users' position, and suggest which characteristics the software should have in order to reduce the chance of being rejected. In the specific case of VJ software, the emphasis on the artistic nature of the live performers' work provides them with a principled safeguard against the very possibility of being effectively replaced by machines, as well as a criterion to evaluate their work. The recommended characteristics of a VJ software would then be: to allow an artist to stop automation and run the software by manual control; to avoid unnecessary complexities in the interface, since technology is needed but the core expertise of the artist lays elsewhere; and to accept clips created with other software, since providing artistic content might eventually compensate VJs for any loss of work as live performers. If these characteristics were not present in the software, DJs and VJs could have the impression of being offered some tool for beginners with poor potential for supporting artistic work, the sophistication of its algorithms notwithstanding.

VJs and DJS use technology as part and parcel of their activity; technology is the means through which they demonstrate their value and gain a competitive advantage over other artists. At the same time, there is a risk that such technology slips out of their control and produces results of apparently equal or higher quality than humans' and that automated VJ software replaces human VJs in venues of very low budget or when the DJ prefers to keep close control on all artistic aspects of their show. Hence, interviewees' arguments seem to echo classic fears raised by automation such as being replaced by machines in the workplace or being surpassed by machines in processes that are considered peculiarly human. However, interviewees avoid simplistic positions, which are so common with technical innovations (Arminen et al., 2016), and which might have emerged had we asked them to rate automation on a scale. When talking spontaneously in their own words, they searched for an adequate use of automation instead of fighting against it, and to identify a threshold to make automation sustainable. Furthermore, having a discretely long expertise in the sector, they could see innovation in the long run, trusting to grow able to develop new ways of expressing themselves and be appreciated. Curiously, some of their arguments are present also in the academic debate on automation: the emphasis on originality reminds of Boden's characterization of creativity as novelty and value (1990), while the emphasis on the human component as necessary to art reminds of Hofstadter's claim that "Art involves the expression and communication of human experience, so that if we did decide that it is the computer which is generating the 'artwork', then it cannot be an art work after all" (Hofstadter, 2002).

There is much debate about the epistemic role of interviews in qualitative research; it is said that they cannot disclose facts, but opinions, that they do not let previously held opinions emerge but instead create new ones, and that they can never avoid the influence of social desirability (Hammersley, 2003; Law, 2009; Schwartz, 2007; Suchman and Jordan, 1990, Whitaker and Atkinson, 2019). Even so, and actually thanks to this, interviews can represent a great resource if their argumentative nature is properly acknowledged. Following the branching patterns of arguments and counter-arguments, the researcher can identify the aspects that would flip users' position from favorable to unfavorable. By tracking down the multiple aspects that are at stake when making a device part of one's professional routine and by considering the views of different and sometimes conflicting stakeholders, the mercurial nature of users' opinion becomes less of a conundrum. The outlining of the argumentative context framing the interviewees' opinions about a new medium, prefaced if necessary by an analysis of the larger background in which this medium lands, is a way to connect such opinions to the cultural and social environment in which new media are supposed to be used: to the consequences of their adoption on the users' community, to the extent to which the quality criteria of a profession are matched and to the perceived effect on the users' social identity. This is why we think that studying users' arguments would help cover the social and cultural aspects of media acceptance that are reputed as important theoretically but have hardly been tackled in the research practice (Bagozzi, 2007).

**Acknowledgments**